\documentclass[aps,twocolumn,prd,showpacs,showkeys,preprintnumbers,superscriptaddress,nobibnotes,floatfix,longbibliography]{revtex4-1}

\usepackage{graphicx}
\usepackage{bm}
\usepackage{times}
\usepackage{hyperref}
\usepackage{slashed}
\usepackage{color}
\usepackage{aas_macros}
\usepackage{nicefrac}
\usepackage{soul}

\usepackage{slashed}
\usepackage{lipsum}
\usepackage{subfigure}
\usepackage{multirow}
\usepackage{amsmath}
\usepackage{array}    
\usepackage{varwidth}
\usepackage{comment}

\graphicspath{{figures/}}
\newcommand{\orcid}[1]
{\begingroup
  \hypersetup{hidelinks}\href{https://orcid.org/#1}{\includegraphics[width=9pt]{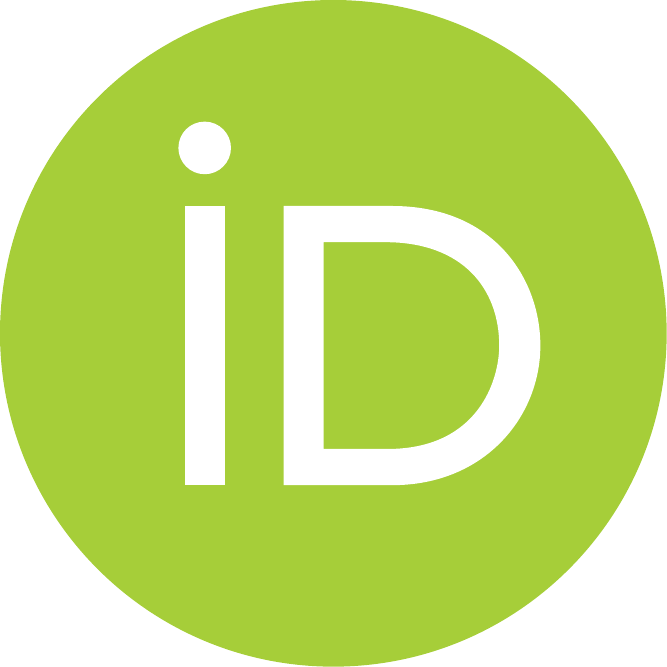}
} \endgroup}

\hypersetup{
    pdfnewwindow=true,    
    colorlinks=true,      
    linkcolor=blue,       
    citecolor=blue,       
    filecolor=blue,      
    urlcolor=blue         
}
\usepackage{xcolor}

\begin{document}

\title{Identifying Extended PeVatron Sources via Neutrino Shower Detection}

\author{Takahiro Sudoh \orcid{0000-0002-6884-1733}}
\email{takahiro\_sudoh@icloud.com}
\affiliation{Center for Cosmology and AstroParticle Physics (CCAPP), Ohio State University, Columbus, OH 43210, USA}
\affiliation{Department of Physics, Ohio State University, Columbus, OH 43210, USA}
\affiliation{Department of Astronomy, Ohio State University, Columbus, OH 43210, USA}
\affiliation{Graduate School of Artificial Intelligence and Science, Rikkyo University, Nishi-Ikebukuro 3-34-1, Toshima-ku, Tokyo 171-8501, Japan}

\author{John F. Beacom \orcid{0000-0002-0005-2631}}
\email{beacom.7@osu.edu}
\affiliation{Center for Cosmology and AstroParticle Physics (CCAPP), Ohio State University, Columbus, OH 43210, USA}
\affiliation{Department of Physics, Ohio State University, Columbus, OH 43210, USA}
\affiliation{Department of Astronomy, Ohio State University, Columbus, OH 43210, USA}

\date{\today}


\begin{abstract}
Identifying the Milky Way's very high energy hadronic cosmic-ray accelerators --- the PeVatrons --- is a critical problem.  While gamma-ray observations reveal promising candidate sources, neutrino detection is needed for certainty, and this has not yet been successful.  Why not?  There are several possibilities, as we delineated in a recent paper \href{https://journals.aps.org/prd/abstract/10.1103/PhysRevD.107.043002}{[T.~Sudoh and J.~F.~Beacom, Phys.~Rev.~D {\bf 107}, 043002 (2023)]}.  Here we further explore the possibility that the challenges arise because PeVatrons have a large angular extent, either due to cosmic-ray propagation effects or due to clusters of sources. We show that while extended neutrino sources could be missed in the commonly used muon-track channel, they could be discovered in the all-flavor shower channel, which has a lower atmospheric-neutrino background flux per solid angle.  Intrinsically, showers are quite directional and would appear so in water-based detectors like the future KM3NeT, even though they are presently badly smeared by light scattering in ice-based detectors like IceCube.  Our results motivate new shower-based searches as part of the comprehensive approach to identifying the Milky Way's hadronic PeVatrons.
\end{abstract}

\maketitle


\section{Introduction}

The Milky Way has powerful but unidentified sources of hadronic cosmic rays (CRs), but their sources are obscured by CR deflections in interstellar magnetic fields~\cite{1964ocr..book.....G, 1990acr..book.....B, 1990cup..book.....G, 2016crpp.book.....G, 2019IJMPD..2830022G}.  Finding and understanding these natural particle accelerators, especially those reaching PeV energies (``PeVatrons"), has been a major problem~\cite{2013A&ARv..21...70B, 2014JHEAp...1....1A, 2021Univ....7..324C}. Candidate hadronic sources have been identified though the emission of TeV--PeV gamma rays, but it is difficult to isolate these from leptonic sources, where the gamma rays are produced by CR electrons~\cite{1971NASSP.249.....S, 2004vhec.book.....A, 2009herb.book.....D}.  Detecting neutrinos from a source would conclusively identify it as hadronic, but this has not yet been successful~\cite{2017PhRvD..96h2001A, 2019ApJ...886...12A, 2020PhRvL.124e1103A, 2020ApJ...892...92A}. In a recent paper~\cite{2023PhRvD.107d3002S}, we use multi-messenger data to constrain general PeVatron models for Milky Way sources.

Milky Way gamma-ray observations reveal a great diversity of types of emission, from point sources to extended sources to diffuse emission from the whole plane~\cite{2007ApJ...664L..91A, 2017ApJ...843...40A, 2018A&A...612A...1H, 2020ApJS..247...33A, 2021Natur.594...33C, 2021PhRvL.126n1101A, 2022arXiv221115321M, 2023RNAAS...7....6A, 2023arXiv230505372C}.  Imaging air-Cherenkov telescopes (IACTs) are especially effective at finding point sources, due to their excellent flux sensitivity and angular resolution, which also makes them superior for determining the morphology of some extended sources. Observatories based on the direct detection of shower particles --- such as water Cherenkov detectors (WCDs) --- are especially effective at finding extended  sources (and diffuse emission), due to their much larger fields of view and exposures.

\begin{figure}[b]
    \centering
    \vspace{-0.1cm}
    \includegraphics[width=\columnwidth]{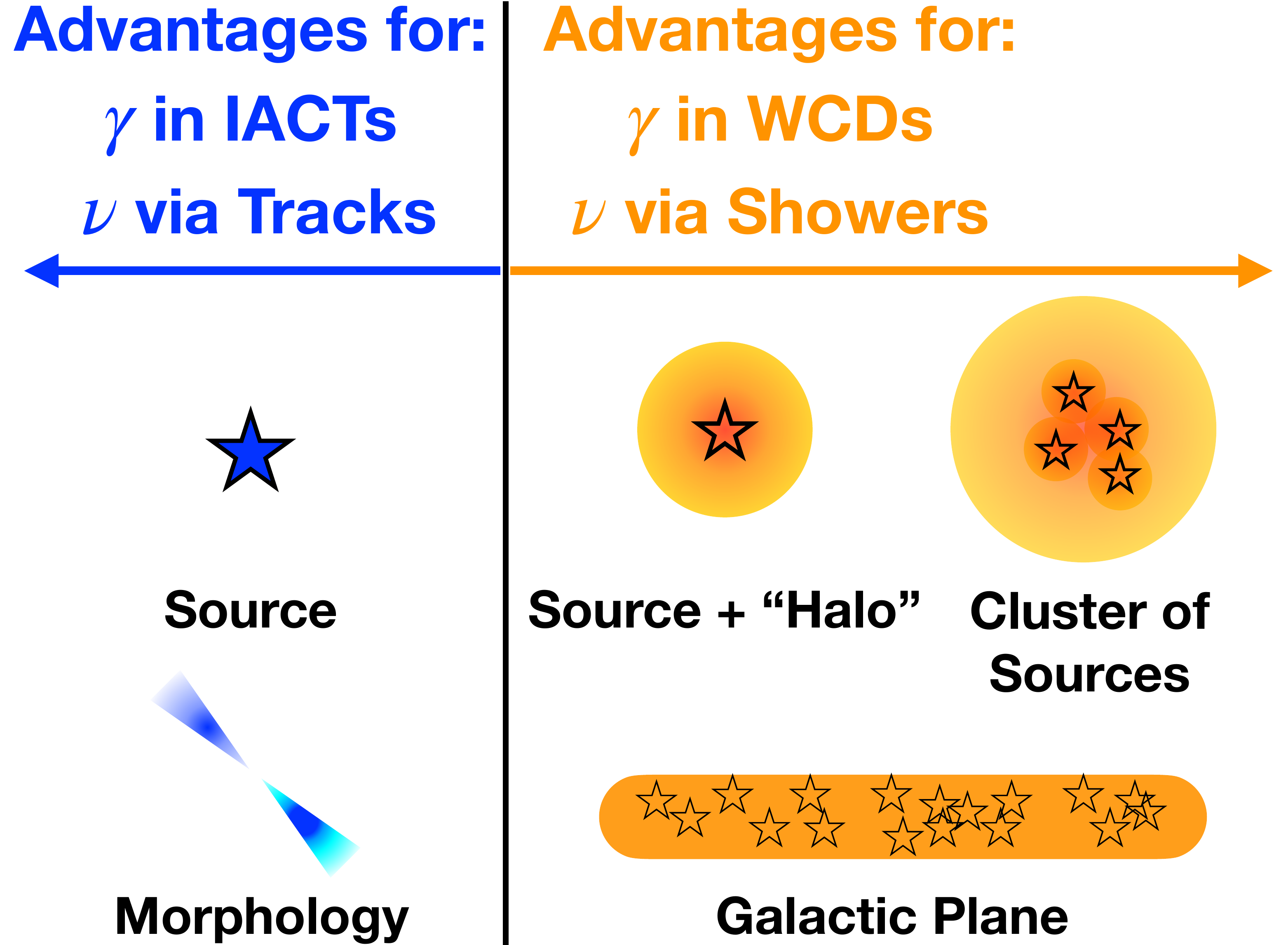}
    \caption{Complementarity of approaches to probing different high-energy gamma-ray and neutrino sources in the Milky Way.  See text.}
    \label{fig:schematic}
\end{figure}

In this paper, we show that a similar complementarity occurs for neutrino source searches with \emph{tracks} versus \emph{showers}. Figure~\ref{fig:schematic} gives a schematic overview.  Tracks, which arise from the charged-current (CC) interactions of muon neutrinos, have good angular resolution and thus have been a major focus in source studies~\cite{2006PhRvD..74f3007K, 2006NIMPA.567..405L, 2006APh....26..310V, 2007PhRvD..75h3001B, 2007ApJ...656..870K, 2009APh....31..376M, 2011APh....34..778V, 2014APh....57...39G, 2017ApJ...836..233G, 2017EPJC...77...66C, 2018APh...100...69A, 2021PhRvD.103j3020N, 2021PhRvD.104b3017N}. Showers, which result from the CC interactions of electron and tau neutrinos, plus the neutral-current (NC) interactions of all flavors, have been much less considered~\cite{2006PhRvD..74f3007K, 2016JPhG...43h4001A, 2019ApJ...886...12A} because they have poor angular resolution in IceCube.  However, this is not an intrinsic limitation --- it is caused by light scattering in ice~\cite{2006JGRD..11113203A} --- and future water-based detectors should do much better, as suggested by the encouraging performance of ANTARES~\cite{2017AJ....154..275A}.  \emph{We show that for a variety of realistic source and future detector scenarios, while the track channel is superior for point sources, the shower channel can be superior for extended sources.}  Shower-based source searches should become an important part of the broader quest to identify hadronic PeVatrons.

In Sec.~\ref{sec:pevatrons}, we review the astrophysical evidence for extended sources.  In Sec.~\ref{sec:method}, we describe our methods to calculate neutrino event spectra. In Sec.~\ref{sec:shower}, we show potential of showers in the general case, while in Sec.~\ref{sec:future}, we focus on near-future specifics. In Sec.~\ref{sec:conclusion}, we conclude.


\section{Prospective Extended PeVatron Sources}
\label{sec:pevatrons}

The nature of the Milky Way's PeVatrons is unknown, despite extensive work.  As a key step forward, in Ref.~\cite{2023PhRvD.107d3002S} we established a new multi-messenger framework, calculating the allowed properties of PeVatrons in light of the observed global CR power budget, TeV--PeV gamma-ray source observations, and TeV--PeV neutrino source limits.  In this section, we review that general framework and also examples of types of extended sources.


\subsection{CR Sources as Neutrino Sources}
\label{subsec:spectrum}

In hadronic accelerators, CR primaries produce gamma-ray and neutrino secondaries through interactions with matter~\cite{1971NASSP.249.....S, 2004vhec.book.....A, 2009herb.book.....D}.  As detailed in Ref.~\cite{2023PhRvD.107d3002S}, there are other important types of high-energy sources: CR electron accelerators that produce only gamma rays (leptonic sources), sources that do not reach PeV energies, sources that are hadronic but are not very luminous, and possible exotic sources (e.g., due to dark matter). Population studies will be needed to categorize how much specific types of gamma-ray sources contribute to the hadronic CR power budget.  Below, we focus on hadronic PeVatrons that contribute appreciably to this budget, but note that our considerations for shower versus track detection apply to any hadronic sources.

Newly produced CRs are temporarily confined near their accelerators by magnetic fields, interacting with matter nearby, thus producing gamma-ray and neutrino emission.  In Ref.~\cite{2023PhRvD.107d3002S}, we therefore characterized potential PeVatrons by the source gas density $n_{\rm src}^{\rm gas}$ and the local confinement time $\tau_{\rm src}^{\rm esc}$.  In the plane of these parameters, we showed the regions allowed by data, plus rough indicators for different types of models.  We showed that while gamma-ray observations are constraining, IceCube and other neutrino detectors have not yet reached the required sensitivity to detect PeVatrons.

To generally characterize PeVatrons, we assume that source protons are produced with a spectrum of the form $dN_p/dE_p \propto E^{-\gamma_p}\exp[-(E_p/E_p^{\rm cut})]$.  Observations indicate that sources inject protons into the interstellar medium with an index around $\gamma_p=2.37$. This is obtained from the index of the observed flux, roughly $E_p^{-2.70}$~\cite{2020APh...12002441L}, by subtracting the assumed energy-dependent index of the escape time, $E_p^{-0.33}$~\cite{2016PhRvL.117w1102A}. For the maximum energy, we assume $E_p^{\rm cut} = 3~\rm PeV$, i.e., the ``knee."~\cite{2008ApJ...678.1165A, 2013arXiv1306.6283T, 2019PhRvD.100h2002A, 2020APh...12002441L} Individual sources may have different spectral indices and cutoffs than assumed here, and even the overall values may be different, but our main results are not sensitive to moderate changes.

The normalization of the proton spectrum is inferred from the CR power budget (for further details, see also Ref.~\cite{2019PhRvD..99f3012M}). For a collection of sources, this is $L_p \sim 1.3 \times 10^{38}$~erg~s$^{-1}$ near 1~PeV (i.e., the differential flux at 1~PeV; here and throughout, when saying \emph{near} we mean integrated over $\Delta\ln E \sim 1$) . At PeV energies, helium and other nuclei become important components of the CR flux~\cite{2022PhRvD.105f3021A}; in Eq.~(\ref{eq:neutrino}) below, we include the enhancement of neutrino emission due to nuclei in the CRs and target medium, following Refs.~\cite{2022A&A...659A..57P, 2023PhRvD.107d3002S}. If CRs are injected impulsively with a source rate of $\Gamma_{\rm CR}$, the CR energy per source is $L_{\rm CR}/\Gamma_{\rm CR}$. We use $\Gamma_{\rm CR} = 0.03$~yr$^{-1}$, i.e., the Galactic core-collapse supernova rate~\cite{2013ApJ...778..164A}, which yields a proton energy per source (integrated above 1~GeV) of $6\times 10^{49}$~erg. Smaller source rates would proportionally increase the luminosity of each source but decrease the number of sources in the Milky Way~\cite{2023PhRvD.107d3002S}.

Given the source proton spectrum, the neutrino spectrum follows $dN_\nu/dE_\nu \propto E^{-\gamma_\nu}\exp[-(E_\nu/E_{\nu}^{\rm cut})^{0.5}]$ with $\gamma_\nu = \gamma_p - 0.1$ (the correction is due to the energy dependence in the \emph{pp} cross section) and $E_{\nu}^{\rm cut} \sim 0.05E_p^{\rm cut}$; for greater precision on the details, see Refs.~\cite{2006PhRvD..74c4018K, 2014PhRvD..90l3014K, 2020ApJ...903...61C}. (The corresponding gamma rays are produced at $E_\gamma \sim 2 E_\nu$.) The neutrino luminosity per source depends on the proton energy budget and the source gas density, $n_{\rm src}^{\rm gas}$. (The confinement time $\tau_{\rm src}^{\rm esc}$ sets the source number~\cite{2023PhRvD.107d3002S}.) Using the above proton spectrum, we obtain the neutrino luminosity near 50~TeV (per-flavor and adding $\nu+\bar{\nu}$):
\begin{equation}
\begin{split}
E_\nu^2 \frac{d^2N_{\nu}}{dE_\nu dt}  &= 6\times 10^{32}~{\rm erg~s^{-1}}~\left(\frac{n_{\rm src}^{\rm gas}}{\rm 10~cm^{-3}}\right)\\
&\left(\frac{E_\nu}{\rm 50~TeV}\right)^{-0.27}\exp{\left(-\sqrt{\frac{E_\nu}{\rm 150~TeV}}\right)},
\label{eq:neutrino}
\end{split}
\end{equation}
where we use $\sigma_{pp} = 60$~mb for $E_p=1$~PeV.  For a reference, this would give a flux of 3$\times$10$^{-12}$~TeV~cm$^{-2}$~s$^{-1}$ at 1~kpc. If such neutrino sources were point-like and located at $\sin(\delta)\gtrsim0$, they should have already been observed by the 10-year IceCube survey at distances up to $\sim$2~kpc~\cite{2020PhRvL.124e1103A}. 


\subsection{Examples of Extended PeVatron Sources}
\label{subsec:size}

Hadronic PeVatrons may have escaped detection in neutrinos because of extended sizes that make them hard to detect via the muon track channel.  From gamma-ray observations, especially with WCDs, we know that there are sources with angular size of a few to several degrees (throughout, all angular sizes are radii unless otherwise noted).  These can be individual sources where CR propagation is different from that in the average interstellar medium or clusters of sources where CR propagation or even acceleration effects may occur in a region.


\subsubsection{Individual Sources with Altered CR Propagation}

Hadronic PeVatrons might appear as point-like gamma-ray and neutrino sources.  For example, if the emission is produced by shocks in supernova remnants (SNRs) at an early stage, pulsar wind nebulae, or  star clusters, the size might be as small as $\sim$1--10~pc. For a source distance of $\sim$1~kpc, the angular size is then $\sim$0.05--0.5$^\circ$, often below the angular resolution of TeV--PeV gamma-ray and neutrino observatories.

However, the emission from such small regions should fade quickly, because the accelerators cannot confine PeV particles for a long time. In SNR shocks, the confinement time for PeV particles is likely 1~kyr at most~\cite{2011MNRAS.410.1577O,2019MNRAS.490.4317C} and could be as small as 10~yr~\cite{2013MNRAS.435.1174S}. While we could look for emission at lower energies, the large atmospheric backgrounds make this difficult; we show below (in Sec.~\ref{sec:shower}) that the most efficient energy range is $E_\nu \gtrsim5$~TeV, which is produced by protons of $E_p\gtrsim100$~TeV. This is high enough that a long confinement time is unlikely. If this were the whole story, it could be difficult to identify hadronic PeVatrons.

These difficulties might be overcome if we look instead for extended diffuse emission in the vicinity of CR sources, produced due to PeV hadronic CRs leaving the acceleration site but not yet effectively escaping into the interstellar medium. If a source produces $\sim$10$^{50}$~erg of CRs that diffuse over $\sim$100~pc, it would produce a CR overdensity of $\sim$0.7~eV~cm$^{-3}$. This is comparable to the typical all-energies CR density in the interstellar medium, suggesting that such an overdensity could be identified. In fact, because CR source spectra are harder than the average CR spectrum, the overdensity would be significantly more prominent at high energies.  If PeV CRs isotropically diffuse at the rate observed for the bulk of the Milky Way, the diffusion coefficient at PeV energies would be $D\sim10^{30}$~cm$^2$~s$^{-1}$~\cite{2020PhRvD.101b3013E}, which implies that the time for CRs to diffuse over 100~pc is $\sim$1~kyr. This exceeds the timescale over which SNRs themselves can confine PeV particles, making PeVatron searches more favorable than they might seem at first. (For relevant models, see Refs.~\cite{2007ApJ...665L.131G, 2009MNRAS.396.1629G})

Importantly, the confinement time around hadronic accelerators can be even longer, because the diffusion coefficient in the vicinity of CR sources may be significantly smaller (about 100 times) than the bulk of the Milky Way. This is exemplified by gamma-ray observations of pulsars surrounded by ``TeV halos" --- a new source class discovered by WCDs and challenging to observe with IACTs~\cite{2007ApJ...664L..91A, 2017Sci...358..911A, 2017PhRvD..96j3016L, 2019PhRvD.100d3016S, 2020A&A...636A.113G, 2022NatAs...6..199L, 2022FrASS...922100F, 2022IJMPA..3730011L, 2023arXiv230402631H}.  (For different interpretations, see Refs.~\cite{2019PhRvL.123v1103L, 2021PhRvD.104l3017R}.) The spatial extent of the slow-diffusion regions is unclear, but Fermi gamma-ray observations of Geminga suggest that the size could be over 100~pc~\cite{2019PhRvD.100l3015D}. This is possible if turbulent magnetic fields have small coherent length and dominate over ordered ones~\cite{2018MNRAS.479.4526L}. (Diffusion can also be suppressed by CR-excited instabilities, although it seems unlikely to sustain these over large scales~\cite{2022PhRvD.105l3008M}.)  Diffusing PeV CRs surrounding a source could then be visible as long as 100~kyr or even longer, greatly improving the prospects for detection.  For a source distance of 1~kpc, a size of 100~pc translates into an angular size of 6$^\circ$, which indicates that observations with even poor angular resolution would suffice.

Gamma-ray observations have also revealed other extended sources. An example is HESS J1825-137, which is believed to be powered by a pulsar and extends to $\sim$1.5$^\circ$~\cite{2019A&A...621A.116H}. The total gamma-ray flux from this source is $E_\gamma^2F_\gamma\sim10^{-12}$~TeV~cm$^{-2}$~s$^{-1}$ near $E_\gamma \sim 100$~TeV. While the emission is thought to be leptonic, this shows that a large energy budget of non-thermal electrons can be transported as far as $\sim$100~pc in an energy-independent manner (assuming a distance of 4~kpc) within $\sim$20~kyr. This suggests that similar extended hadronic sources could also exist.  For even larger sources, Ref.~\cite{2022PhRvD.106l3029G} proposed that very extended --- ten degrees and above --- hadronic emission due to  escaping PeV protons over a $\sim$1~kpc scale might be detectable in the map of the diffuse Galactic gamma-ray emission. The estimated flux level reaches $\sim$$10^{-12}$~TeV~cm$^{-2}$~s$^{-1}$. The neutrino counterpart would be an intriguing target for showers (see also Ref.~\cite{2022PhRvD.106f3004B} for another extended-source model).

Last, it is possible that there are exotic sources of large extent.  This could include the annihilation or decay emission from dark-matter subhalos (see, e.g., Refs.~\cite{2012A&A...538A..93Z, 2015JCAP...12..035B, 2016JCAP...05..028S, 2023arXiv230400032B}, which searched for such sources in Fermi gamma-ray data).


\subsubsection{Clusters of Sources}

Gamma-ray observations also reveal extended emission due to collections of sources. Even where single-source detection is not possible, we may observe the \emph{total} emission from sources and nearby diffusing CRs.  In some cases, this may simply be a sum of sources; in others, there may be collective effects that modify CR propagation or acceleration, e.g., in superbubbles~\cite{2020SSRv..216...42B, 2021MNRAS.504.6096M, 2022MNRAS.512.1275V}. While it would be challenging to isolate the source components, such observations would still be important to narrow down the nature of hadronic PeVatrons. 

One important extended target is the Cygnus region, highlighted long ago in Refs.~\cite{2007PhRvD..75h3001B, 2007PhRvD..76f7301A, 2007PhRvD..75f3001A, 2007PhRvD..76l3003H}.  This is an active star-forming region, containing sources like SNRs, pulsar wind nebulae, and star clusters. Diffuse gamma-ray emission from this region is also observed~\cite{2007ApJ...658L..33A}. Toward the Cygnus OB2 association, Fermi and HAWC have observed a ``cocoon" of gamma-ray emission, which extends to $\sim$$2^\circ$ and has a flux of $E_\gamma^2F_\gamma\sim2\times10^{-12}$~TeV~cm$^{-2}$~s$^{-1}$ at 100~TeV~\cite{2021NatAs...5..465A}. In fact, some emission might be present at larger radius, because the CR energy density is still large at $\sim$$2^\circ$. Coincident with this is the observation of the Galactic Plane in sub-PeV gamma rays by Tibet AS$\gamma$, which finds a hint of excess events from a $\sim$4$^\circ$ region toward the Cygnus Cocoon~\cite{2021PhRvL.126n1101A, 2021ApJ...914L...7L, 2021ApJ...919...93F}.  These results suggest that the total source + diffuse neutrino emission on a scale of $\sim$2--4$^\circ$ may be detectable from the Cygnus region. (A hint of such emission has been  reported recently~\footnote{A. Neronov, at {\it Neutrinos in the Multi-Messenger Era}, 2022 \href{https://agenda.irmp.ucl.ac.be/event/4681/}{https://agenda.irmp.ucl.ac.be/event/4681/}}.)

Another important extended target is the Galactic Center region, which is rich in CR accelerators and dense gas clouds, making it an excellent target for neutrino searches. A region of dense gas ($>$100~cm$^{-3}$) called the Central Molecular Zone has a disk-like shape with an angular extent (not the radius) of about $3^\circ\times0.5^\circ$~\cite{2022arXiv220311223H} ($\sim$400~pc$\times$$\sim$70~pc for a distance of $\sim$8~kpc). On larger scales, there is a region of still higher gas density ($\sim$30~cm$^{-3}$) over the angular size of about $7^\circ\times1.7^\circ$ (corresponding to a radius of $\sim$1~kpc and a height of $\sim$250~pc)~\cite{2020ApJ...891..179H, 2006PASJ...58..847N, 2016PASJ...68....5N}. Hadronic CRs can reach (and also may be produced in) such extended regions; assuming $D\sim10^{30}$~cm$^2$~s$^{-1}$ for PeV protons, particles injected 100~kyr ago at the Galactic Center would diffuse out to $\sim$1~$\rm kpc$. While H.E.S.S., MAGIC, and VERITAS has observed diffuse gamma-ray emission at the level of $\sim$$10^{-12}$~TeV~cm$^{-2}$~s$^{-1}$ near 10~TeV that extends over $\sim$$3^\circ$ in longitude and $\sim$$0.5^\circ$ in latitude~\cite{2018A&A...612A...9H, 2020A&A...642A.190M, 2021ApJ...913..115A}, even greater total fluxes may be present over larger angular scales. A hint of neutrino emission is also suggested by recent ANTARES analysis~\cite{2022arXiv221211876A}, which uses a region of $|l|<30^\circ$ and $|b|<2^\circ$.

While not strictly a potential PeVatron, the emission from the Galactic Plane stands as another extended source.  While observing it would provide important inputs toward understanding the production and propagation of hadronic CRs, it cannot by itself reveal the locations of sources. This is because this emission is produced by the interactions of diffusing CRs with interstellar gas plus the contributions from unresolved sources. We do not discuss this component in detail, but note that showers should be of primary importance to study it.


\section{Neutrino event rates and spectra}
\label{sec:method}

In this section, we outline our methods to predict the detected event rates, following the ``theorist's approach" of Refs.~\cite{2006PhRvD..74f3007K, 2013PhRvD..88d3009L, 2017PhRvD..96j3006N} (building on Ref.~\cite{1990cup..book.....G}), which is suitable for discussing general physics points and gives results accurate within $\sim$30\%, which is adequate at present, as discussed below.  Importantly, we calculate rates as a function of \emph{detectable} energy, $E_{\rm det}$, instead of neutrino energy, $E_\nu$, to more realistically compare signals and backgrounds, as the differential cross sections and kinematics vary with the interaction channel.

We seek to estimate the possible flux sensitivities that can be achieved with shower detection under hypothetical but plausible setups.  On one hand, the event rates could be somewhat smaller than presented below due to cuts to remove backgrounds that also reduce signal rates (especially below $E_{\rm det}\sim1$~TeV, though we show below that this energy range is not very important).  On the other hand, the experiments may be able to use improved event-reconstruction techniques to reach better angular resolution than we assume below.  In support of our approach, we show below that we are able to reproduce key results from the KM3NeT collaboration.  In the long term, detailed simulations from the collaborations will be needed to definitively assess sensitivities to extended sources.


\subsection{General Formulae for Event Rates}

Here, we discuss the neutrino event spectra from a given source flux $F_{\nu}$, by which we always mean $\nu + \bar{\nu}$ and integrated over the source size. Backgrounds are characterized by the intensity (flux per solid angle), $\Phi_\nu$, which needs to be integrated over the solid angle, as discussed in Sec.~\ref{subsec:bkg}. After reviewing general points, we introduce shower events first, for which the spectrum calculation is conceptually simpler, and then track events.

The detection of neutrinos is based on the deep inelastic scattering of neutrinos with quarks, which produces observable signals in the detector. We take the neutrino cross sections, $\sigma$ from from Ref.~\cite{1998PhRvD..58i3009G} and inelasticity values, $y$, from Ref.~\cite{1996APh.....5...81G} (below, we attach a CC or NC subscript as needed); Using instead a more recent paper~\cite{2011PhRvD..83k3009C} does not change our results. In a CC interaction, a $W$-boson is exchanged and a neutrino $\nu_l$ is converted to a charged lepton $l$ with energy $(1-y_{\rm \scriptscriptstyle CC})E_\nu$, accompanied by a hadronic shower that carries the remaining energy. In a NC interaction, a $Z$-boson is exchanged and a hadronic shower is produced with energy $y_{\rm \scriptscriptstyle NC}E_\nu$, while the energy in the outgoing neutrino is undetected. The cross sections increase with the neutrino energy; below $\sim$5~TeV, these are nearly proportional to $E_\nu$; at higher energies, the dependence approaches $E_\nu^{0.4}$. In the energy range of interest, $\sigma_{\rm \scriptscriptstyle NC}$ is smaller than $\sigma_{\rm \scriptscriptstyle CC}$ by a factor of $\sim$0.4. The distribution of inelasticity, $d\sigma/dy$, peaks at $y=0$ and has an average value of $\langle y \rangle \sim 0.4$ for both the CC and NC cases in the energy range of interest. (The value of $\langle y \rangle$ is larger for neutrinos than for antineutrinos, but we average the values.)

Shower (or ``cascade") events are produced largely by the CC interactions of electron and tau neutrinos.  (Muon-neutrino CC events are efficiently separated from showers due to the long muon tracks.)  In the electron-neutrino case, the full neutrino energy is transferred to the combined hadronic and electromagnetic shower, resulting in $E_{\rm det}\simeq E_\nu$.  In the tau-neutrino case, this is nearly true at these energies because the tau lepton decays promptly, though here are $\sim$30\% losses due to the decay energy carried by neutrinos~(Chapter 58, Volume 1 in Ref.~\cite{ParticleDataGroup:2022pth}), which affects our total CC shower event-rate calculation at the $\sim$20\% level. We also ignore the fact that $\sim$17\% of $\nu_\tau$ CC events produce muon tracks through tau decays because this affects our total calculation at less than 10\%. We also ignore tau re-generation effects, which are small, especially in this energy range, because Earth attenuation is modest below about 40 TeV~\cite{2002PhRvD..66b1302B, 2022arXiv220313827A}.   The detectable spectrum is then
\begin{equation}
\begin{split}
   \frac{dN_{\rm sh}}{dE_{\rm det}} &= {T V N_A\rho_{\rm det}} \\
   &\times \sigma_{\rm \scriptscriptstyle CC}(E_\nu) F_{\nu_e+\nu_\tau}(E_\nu) e^{-\tau_\oplus(E_\nu)}\biggr\rvert_{E_\nu = E_{\rm det}},
\end{split}
\end{equation}
where $V$ is the detector physical volume and $\rho_{\rm det}$ is its density, $T$ is the observation time, $N_A$ is the Avogadro number (in units of nucleons per gram), $\tau_\oplus$ is the Earth optical depth in a particular direction, and we sum over $\nu_e$ and $\nu_\tau$ in the flux. We use $\rho_{\rm det} = 1$~g~cm$^{-3}$ (and would use 0.9~g~cm$^{-3}$ for ice).

The NC interactions of all flavors also produce showers through the energy deposited in hadronic showers of energy $E_{\rm det}\simeq \langle y_{\rm \scriptscriptstyle NC}\rangle E_\nu$. The spectrum is 
\begin{equation}
\begin{split}
   \frac{dN_{\rm sh}}{dE_{\rm det}} &= \frac{T V N_A\rho_{\rm det}}{\langle y_{\rm \scriptscriptstyle NC}\rangle} \\
   &\times \sigma_{\rm \scriptscriptstyle NC}(E_\nu) F_{\nu_e+\nu_\mu+\nu_\tau}(E_\nu) e^{-\tau_\oplus(E_\nu)}\biggr\rvert_{E_\nu = \frac{E_{\rm det}}{\langle y_{\rm \scriptscriptstyle NC}\rangle }}.
\end{split}
\end{equation}
NC showers are suppressed in importance relative to CC showers due to two factors. First, $\sigma_{\rm \scriptscriptstyle NC}$ is smaller than $\sigma_{\rm \scriptscriptstyle CC}$, though six flavors contribute instead of four. Second, the energy deposition is smaller, which is important given the steeply falling neutrino spectra. In combination, for a spectrum scaling as $F_\nu \propto E_\nu^{-\gamma}$ and a cross-section scaling as $\sigma \propto E_\nu^\omega$, the total NC spectrum as a function of $E_{\rm det}$ is suppressed compared to the total CC spectrum by factor $\sim$$(6/4)(\sigma_{\rm \scriptscriptstyle NC}/\sigma_{\rm \scriptscriptstyle CC})  \langle y_{\rm \scriptscriptstyle NC} \rangle^{\gamma - 1 - \omega}$, which is $\sim$1/5 for assumed sources ($\gamma\simeq2.3$) and $\sim$1/20 for the atmospheric background ($\gamma\simeq3.7$). Despite this, the background from NC $\nu_\mu$ showers is still important compared to CC $\nu_e$ showers, because the atmospheric $\nu_\mu$ flux is $\sim$20 times larger than the $\nu_e$ flux.

Tracks are created by two types of events. When CC interactions of muon neutrinos with nucleons take place inside the detector (a ``contained-vertex" event), the spectrum is 
\begin{equation}
\begin{split}
   \frac{dN_\mu}{dE_{\rm det}} &= \frac{T V N_A \rho_{\rm det}}{1-\langle y_{\rm \scriptscriptstyle CC}\rangle(E_\nu)} \\
   &\times \sigma_{\rm \scriptscriptstyle CC}(E_\nu)F_{\nu_\mu}(E_\nu) e^{-\tau_\oplus(E_\nu)}\biggr\rvert_{E_\nu  = \frac{E_{\rm det}}{1-\langle y_{\rm \scriptscriptstyle CC}\rangle}},
\end{split}
\end{equation}
where we choose to set the detectable energy as the energy of the muon when it is {\it created}: $E_{\rm det}\simeq(1-\langle y_{\rm \scriptscriptstyle CC}\rangle)E_\nu$. If the energy in hadronic showers can also be estimated, we can instead set $E_{\rm det}\simeq E_\nu$, close to what is done by IceCube~\cite{2021PhRvD.104b2002A}. While both choices are reasonable, ours gives a slightly (by $\sim$10\%) better flux sensitivity from track events, because we assume a hard source spectrum on top of the steeply falling background. (If the source spectrum is soft, our choice might give slightly worse flux sensitivity.)

Muon-neutrino interactions can also be detectable  when these take place outside the detector (a ``through-going" event). The muon spectrum is
\begin{equation}
\begin{split}
    \frac{dN_\mu}{dE_{\rm det}} = &\frac{T A N_A}{\alpha_\mu + \beta_\mu E_{\rm det}} \\
    &\times\int_{\frac{E_{\rm det}}{1-\langle y_{\rm \scriptscriptstyle CC}\rangle}}^\infty dE_\nu \, \sigma_{\rm \scriptscriptstyle CC}(E_\nu)F_{\nu_\mu}(E_\nu) e^{-\tau_\oplus(E_\nu)},
\end{split}
\label{eq:ana_going_main}
\end{equation}
where $E_{\rm det}$ corresponds to the muon energy when it \emph{enters} the detector and $A$ is the detector physical area. We have verified with numerical calculations that integrating over the full $y_{\rm \scriptscriptstyle CC}$ distribution and considering the stochastic nature of muon energy losses changes our results by less than $\sim$20$\%$ (see Appendix~\ref{app:throughgoing}). The properties of the material outside the detector where the muons are born and propagate is included in the energy-loss parameters $\alpha_\mu$ and $\beta_\mu$. We use $\alpha_\mu = 3 \times 10^{-6}$~TeV~cm$^2$~g$^{-1}$ and $\beta_\mu = 3 \times 10^{-6}$~cm$^2$~g$^{-1}$, which are obtained for water~\cite{Groom:2001kq}; choosing instead standard rock would reduce the muon event rate by $\sim$30\% (see Appendix~\ref{app:throughgoing}).  As through-going muons would be produced partially in water and partially in rock, the net effect is within our precision goals.


\subsection{Angular Resolution for Tracks and Showers}
\label{subsec:shower-ang}

Shower events do not have good angular resolution in IceCube ($\sim$20$^\circ$~\cite{2019ApJ...886...12A}) mostly due to Cherenkov light scattering in ice.  The resolution can be significantly improved for water-based detectors; ANTARES demonstrates that they are able to reach $\sim$2--3$^\circ$~\cite{2017AJ....154..275A} at relevant energies, which can be further improved in future telescopes like KM3NeT. The angular resolution that could \emph{ideally} be achieved is much better.  In CC interactions, which dominate the event rate, most of the energy initially goes into an electromagnetic shower.  Electromagnetic showers are $\sim$10~m long, with a transverse spread of $\sim$0.1~m~\cite{2020pprl.book.....F}; these numbers set a scale for the angular resolution of $\sigma_{\rm ang}^{\rm sh}\sim0.1/10\sim0.6^\circ$.  In NC interactions, which are less important, most of the energy initially goes into the hadronic shower.  Hadronic showers have a similar length but a wider transverse size of $\sim$1~m, which seems less favorable.  However, initially hadronic showers induce electromagnetic showers, and the higher the shower energy, the larger the fraction of the Cherenkov light that is produced by the electromagnetic component~\cite{Li:2016kra}.

In the following, we use $\sigma_{\rm ang}^{\rm sh} = 2^\circ$ as a fiducial scale, and also use smaller values down to $0.6^\circ$ to show how improving the shower angular resolution would help. This fiducial angular scale is larger than that of typical gamma-ray sources ($\sim$0.3$^\circ$), so improving the resolution would significantly reduce atmospheric-neutrino backgrounds.

Muon track events have excellent angular resolution. At high energies, $E_\nu\gtrsim10$~TeV, this could be as good as $\sim$0.4$^\circ$ for IceCube~\cite{2020PhRvL.124e1103A} and $\sim$0.1$^\circ$ for KM3NeT/ARCA~\cite{2016JPhG...43h4001A} (the latter only $\sim$2 times worse than might be expected from a deflection of $\sim$1~m over 1 km).  At lower energies, the muon angular resolution is limited by the kinematic angle between neutrinos and muons in CC interactions, $\sigma_{\rm ang}^{\mu, \rm kin}$, which is irreducible unless the muon track and the hadronic shower could be separately reconstructed.  At higher energies, the dominant component is the reconstruction error of the muon direction, $\sigma_{\rm ang}^{\mu, 0}$, which depends on the detector specifics. We thus take the muon angular resolution to be
\begin{equation}
    \sigma_{\rm ang}^{\mu} = \sqrt{(\sigma_{\rm ang}^{\mu, \rm kin})^2 + (\sigma_{\rm ang}^{\mu, 0})^2}.
\end{equation}
In the following, we use $\sigma_{\rm ang}^{\mu, \rm kin}=0.5^\circ(E_\nu/\rm TeV)^{-0.5}$ and $\sigma_{\rm ang}^{\mu, 0} = 0.1^\circ$ (which could be appropriate for water-based detectors) as a fiducial scale, and briefly discuss the case of $\sigma_{\rm ang}^{\mu, 0}=0.4^\circ$.


\subsection{Atmospheric Muon and Neutrino Backgrounds}
\label{subsec:bkg}

The interactions of CRs with nuclei in the atmosphere result in huge backgrounds of muons and neutrinos. We only consider sources below the horizon, which removes atmospheric muons due to the absorption by Earth.  {However, misreconstructed muons and the showers they induce can cause backgrounds that mimic shower and track events with a variety of apparent directions.  Careful cuts are needed to remove these without reducing the signal efficiency too much.} 

The major contribution to the atmospheric-neutrino flux is from the decay of charged pions and kaons (the ``conventional" component), for which we use the neutrino fluxes calculated by Honda {\it et al.}~\cite{2015PhRvD..92b3004H}. Specifically, we use their fluxes calculated at solar minimum averaged over azimuth angle and in zenith angle bin of $\Delta\cos\theta_Z=0.1$; the approximations corresponding to these choices are very small.  In addition, we use their results for a South Pole location, but this choice barely affects our results. Above 10 TeV, where the results of Ref.~\cite{2015PhRvD..92b3004H} are not available, we extrapolate assuming that the spectrum follows $E^{-(\gamma_{\rm CR}+1)}$, where $\gamma_{\rm CR}$ is the index of the parent CR spectrum and the steeper neutrino spectrum is due to pion energy loss in the atmosphere~\cite{2002ARNPS..52..153G}.  We choose the conventional value of $\gamma_{\rm CR}=2.7$ and do not include the steepening at the knee.  A minor contribution to the atmospheric-neutrino flux is due to the decays of short-lived charmed mesons (the ``prompt" component), which produce an near-isotropic spectrum that follows the cosmic-ray spectrum.  For this, we use the model of Enberg {\it et al.}~\cite{2008PhRvD..78d3005E}. Our conclusions are unchanged if we use more detailed calculations based on Refs.~\cite{2015EPJWC..9908001F, Heck:1998vt, 2002JGRA..107.1468P, Riehn:2017mfm}.

Importantly, \emph{the shower channel has a nearly an order-of-magnitude lower atmospheric-neutrino background flux per solid angle than the track channel}. A major component of the background in the shower channel comes from the CC interactions of atmospheric $\nu_e$, for which the flux is by more than an order of magnitude less than for $\nu_\mu $~\cite{2004JCAP...11..009B}. Another major component comes from the NC interactions of $\nu_\mu$, but this is reduced compared to the CC component, as discussed above. 

The diffuse astrophysical neutrino flux is an additional background. We model this component by using the measurements of Ref.~\cite{2020PhRvL.125l1104A}, which obtained a best-fit power-law of $E_\nu^2\Phi_\nu = 1.7\times 10^{-11} (E_\nu/\rm 100~TeV)^{-0.53}$~TeV~cm$^{-2}$~s$^{-1}$~sr$^{-1}$ (per flavor). This is primarily extragalactic in origin, with the Milky Way contribution limited to less than $\sim$10$\%$~\cite{2017ApJ...849...67A, 2018ApJ...868L..20A}.  This power-law function is determined based on the shower analyses that are best optimized for $E_\nu\gtrsim16$~TeV. Because the diffuse cosmological neutrino flux may have a low-energy cutoff, extrapolating a single power law to low energies might significantly overestimate the background below $\sim$10~TeV (see, e.g., Ref.~\cite{2016PhRvL.116g1101M}). This would lead to an underestimate of the statistical power of shower events, for which astrophysical backgrounds are important, but not tracks, for which the atmospheric-neutrino background dominates. Our conclusions are unchanged if we instead use a model for Galactic diffuse neutrino emission in Ref.~\cite{2022arXiv221115607S} as the astrophysical background. 

We obtain the background flux by multiplying the intensity by a solid angle, $\Omega_{\rm bkg}$, which is defined by the source size and detector angular resolution. For the source size, we use the radius containing 68$\%$ of the flux, $\theta_{\rm src}$. For the angular resolution, care is needed because what is often quoted is the standard deviation $\sigma_{\rm ang}$ of two-dimensional Gaussian, an area which contains only 39$\%$ of the total flux from a point source. (Another typical convention is the median angle between simulated neutrinos and reconstructed direction.) We instead use a larger radius, $1.5\sigma_{\rm ang}$, corresponding to the 68$\%$ containment radius. Then, we set $\Omega_{\rm bkg}=\pi[(1.5\sigma_{\rm ang})^2 + (\theta_{\rm src})^2]$. As this solid angle contains only 68$\%$ of the source emission, we reduce the source flux accordingly.

Having calculated the signal ($S$) and background ($B$), we calculate the test statistic (TS) as
\begin{equation}
{\rm TS} = 2 \times [B - N + N\log(N/B)],
\label{eq:test-statistic}
\end{equation}
where $N = S + B$ (see Ref.~\cite{2014APh....57...39G}). The significance can be approximated by $\sqrt{\rm TS}$, which in some limited cases approaches to $S / \sqrt{B}$. We do not take uncertainties in the background into account; this may be needed for more careful assessment of the sensitivity, which is beyond the scope of this paper. We consider targeted searches toward a specific PeVatron candidate and thus do not take trials factors to take into account the ``look-elsewhere effect”~(Chapter 40 in Ref.~\cite{ParticleDataGroup:2022pth}) for observing multiple potential sources.


\section{Shower Versus Track Detection}
\label{sec:shower}

In this section, we calculate the flux sensitivity for extended neutrino sources as observed with showers and tracks, focusing on generic scenarios.

\begin{figure*}
\includegraphics[width=2\columnwidth]{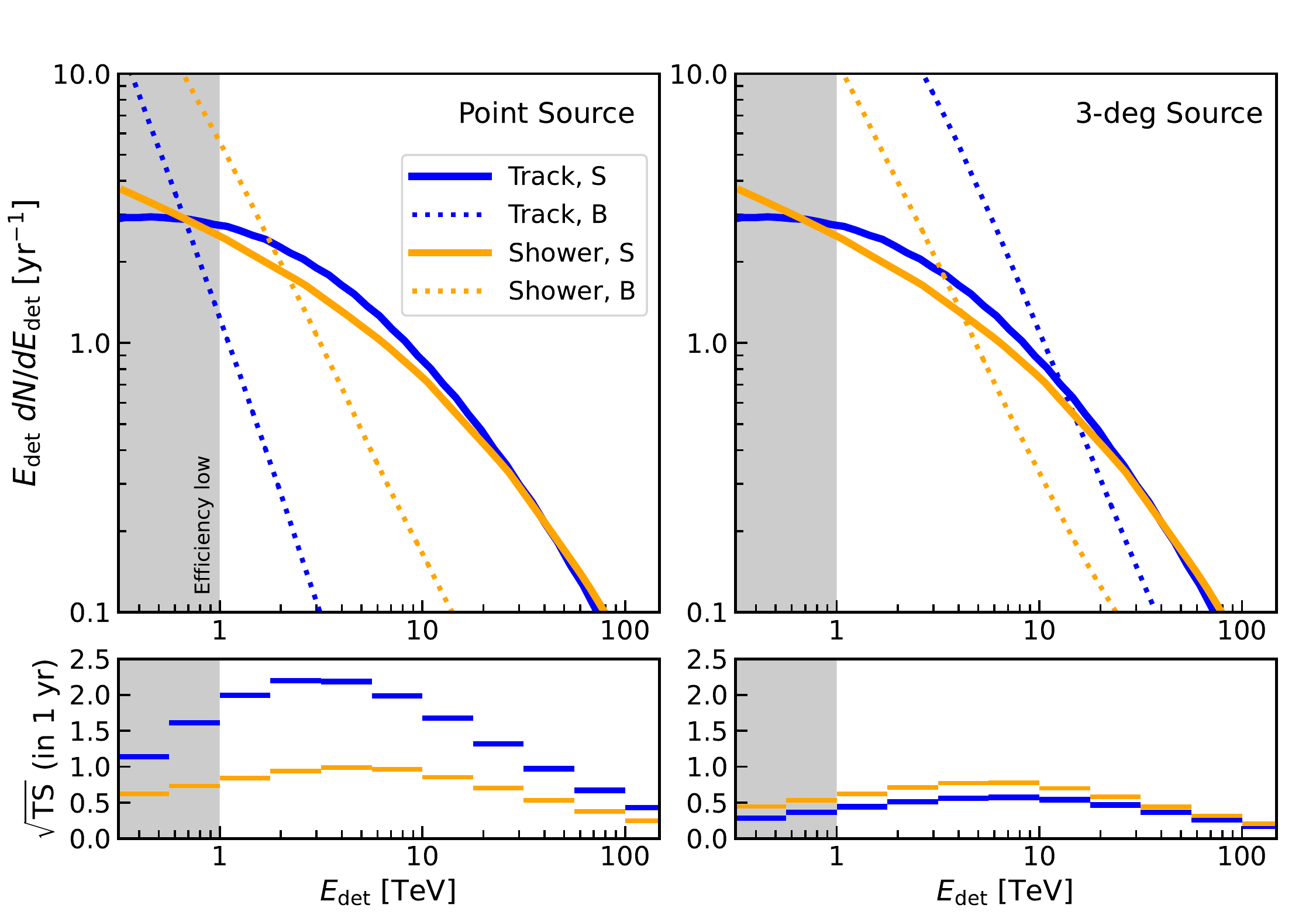}
    \caption{{\bf Upper:} Event rates from the source (solid) and background (dotted) for tracks (blue) and showers (orange).  The 50-TeV flux is fixed to 3$\times$10$^{-12}$~TeV~cm$^{-2}$~s$^{-1}$ and the source neutrino spectrum shape follows $(E_\nu)^{-2.27}\exp(-\sqrt{E_\nu/\rm 150~TeV})$, as per Eq.~(\ref{eq:neutrino}). We assume a km$^3$ detector with angular resolution values of $\sigma_{\rm ang}^{\mu, 0}=0.1^\circ$ and $\sigma_{\rm ang}^{\rm sh}=2^\circ$. {\bf Lower:} The square root of TS (Eq.~\ref{eq:test-statistic}) calculated from $S$ and $B$ within bins of detectable energy. The region $E_{\rm det}<1$~TeV is shaded in gray because the efficiency of realistic detectors most likely drops significantly below this. The source angular radius is set to 0$^\circ$ {\bf(left)} and 3$^\circ$ {\bf (right)}.}
    \label{fig:spectrum}
\end{figure*}

\begin{figure}
    \centering
    \includegraphics[width=\columnwidth]{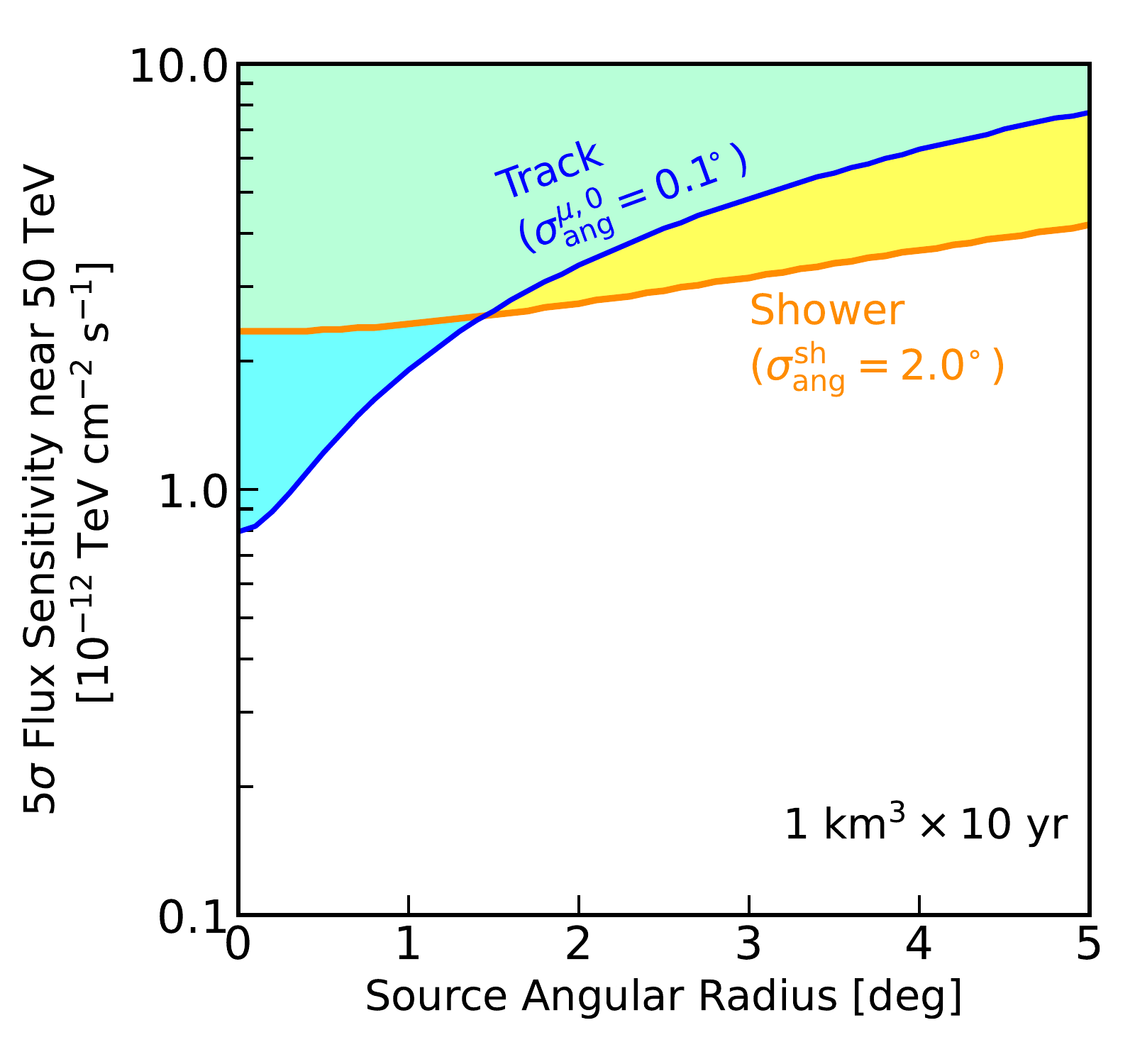}
    \caption{Neutrino flux sensitivities for tracks (blue line) versus showers (orange line). Sources in the cyan and yellow area are detected by tracks and showers respectively; those in green are detected by both. The flux is per flavor (summing $\nu + \bar{\nu}$, as throughout the paper), with the spectrum shape as per Eq.~(\ref{eq:neutrino}). \emph{For extended sources, showers can perform better than tracks.}}
    \label{fig:main}
\end{figure}

\begin{figure}
    \centering
    \includegraphics[width=\columnwidth]{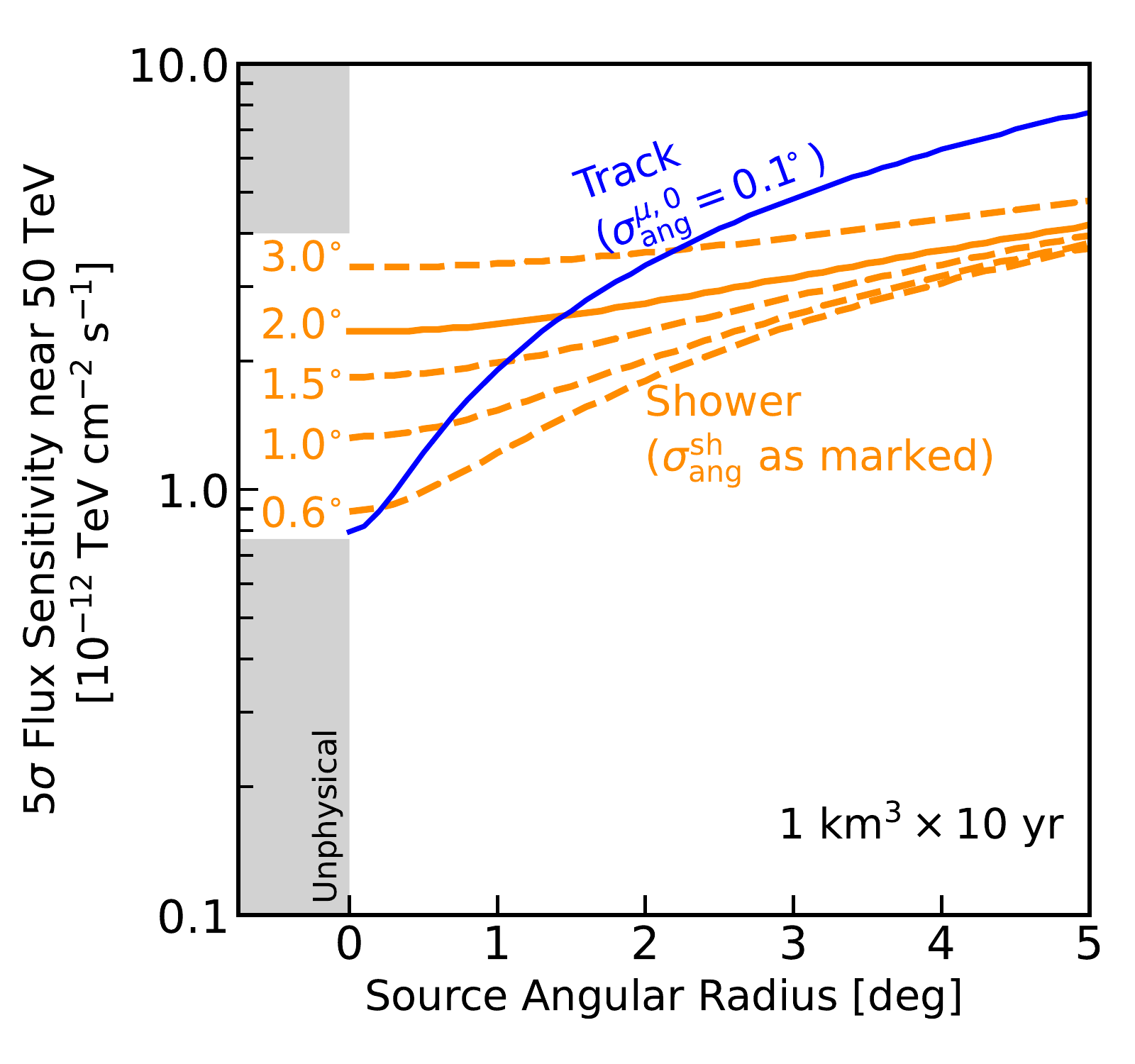}
    \caption{Same as Fig.~\ref{fig:main}, but for various shower angular resolution values.  For clarity, the color shading is omitted. \emph{Improving shower angular resolution has a big impact on source detectability.}}
    \label{fig:main-sh-ang}
\end{figure}

We assume a 1 km$^3$ detector ($A$ = 1 km$^2$ and $V$ = 1 km$^3$).  For reference, we assume it to be located in the Mediterranean Sea (latitude = 39.3$^\circ$).  The main point of this specificity is just that neutrino observatories in different locations see different parts of the Milky Way.  For track-like events, we add the contributions from contained-vertex and through-going muons (using the corresponding definitions of $E_{\rm det}$), though they could be separated. For the energy range of interest, through-going muons dominate.  For shower events, we add contributions from CC and NC interactions (again using the corresponding definitions of $E_{\rm det}$), where the former dominates, as explained above.  The declination of a generic source is set to $-60^\circ$, so that the source is always visible as upgoing for a detector in the Mediterranean. 

Figure~\ref{fig:spectrum} (upper) shows the expected event rates from the signals and backgrounds, considering both the track and shower channels.  We show the event rate as $E_{\rm det}dN/dE_{\rm det}=dN/d\ln E_{\rm det}=(2.3)^{-1}dN/d\log_{10} E_{\rm det}$ as a function of $\log_{10} E_{\rm det}$. The source neutrino flux is normalized to 3$\times$10$^{-12}$~TeV~cm$^{-2}$~s$^{-1}$ near 50~TeV; this corresponds to a distance of 1~kpc for our reference source, as per Eq.~(\ref{eq:neutrino}).  The energy range below 1~TeV is shaded to indicate that the efficiency of realistic detectors drops significantly here.  In any case, we find that the threshold energy $E_{\rm det, Th}$ that maximizes TS is greater than 1~TeV in all cases.

In the left panel, the source size is set to 0$^\circ$ to show that the track channel is much more powerful than the shower channel for observing point sources. In the right panel, the source size is set to 3$^\circ$, as might be expected for extended sources. While both the track and shower channels have comparable signals, showers have smaller backgrounds because of the low intensity of atmospheric $\nu_e $ and the suppression of the NC component. 

Figure~\ref{fig:spectrum} (lower) shows the square root of TS calculated in small bins of detectable energy. The purpose of this plot is to determine the energy range most important for source detection based on the competition of signals and backgrounds relative to the statistical uncertainties.  For both tracks and showers, the most important energies are $E_{\rm det} \sim$ 5~TeV, which provides a target to optimize searches.  These results are calculated for our reference source, following Eq.~\ref{eq:neutrino}, and depend somewhat on assumptions about the spectrum; a softer spectrum would make lower-energy events more important, closer to where the detector efficiency plummets.

To determine if a source can be detected, we use the energy-integrated TS above a threshold energy, $E_{\rm det, Th}$. Specifically, we calculate $S(>E_{\rm det, Th})$ and $B(>E_{\rm det, Th})$ and use Eq.~(\ref{eq:test-statistic}) to calculate the TS. The threshold $E_{\rm det, Th}$ is set such that TS is maximal. With these choices, the three-year 5$\sigma$ flux sensitivities (in $E_\nu^2 F_\nu$) for a point source with an $E^{-2}$ spectrum are $2\times 10^{-12}$ TeV~cm$^{-2}$~s$^{-1}$ for tracks and $8\times 10^{-12}$ TeV~cm$^{-2}$~s$^{-1}$ for showers; these results are consistent with an independent study for KM3NeT~(Fig.~45 in Ref.~\cite{2016JPhG...43h4001A}), which shows that our methods are reasonable.

The hypothetical source in the right panel of Fig.~\ref{fig:spectrum} is detectable in one year at 1.5$\sigma$ with showers and 1.0$\sigma$ with tracks; 5$\sigma$ detection can be expected in $\sim$11~yr for showers and $\sim$24~yr for tracks. For this extended source, showers perform better than tracks.

Figure~\ref{fig:main} illustrates our main focus; this shows the 5$\sigma$ flux sensitivity in 10 years of observations as a function of source size, for showers in comparison with tracks. Other than the source size, the assumptions follow Fig.~\ref{fig:spectrum}.  The y-axis is the flux sensitivity near 50~TeV. The yellow area shows the unique discovery space that showers can probe relative to tracks.   Even with a modest angular resolution of $\sim$2$^\circ$, showers can have better sensitivities than tracks for sources larger than 1$^\circ$. In fact, over of the most parameter space, the flux sensitivities for showers and tracks are comparable (within a factor of $\sim$2), which implies that the combination of these two channels would be important to enhance source detectability even when the shower channel is not more powerful.

Figure \ref{fig:main-sh-ang} shows how the flux sensitivity depends on the shower angular resolution.  Improving shower directional reconstruction could have a large impact on finding neutrino sources. For reference, in Appendix~\ref{app:mu_ang}, we show the case where the track angular resolution is worse ($\sigma_{\rm ang}^{\mu, 0}=0.4^\circ$). Except for point sources, the effect of changing the muon angular resolution is small, which implies that improving this is not important for source detection, though it could be useful for measuring source morphology.

While the advantages of shower detection for extended sources is clear, the results in Fig.~\ref{fig:main} may still look pessimistic; extended emission even as bright as 10$^{-12}$~TeV~cm$^{-2}$~s$^{-1}$ may not be detected at 5$\sigma$ with a 1 km$^3$ detector. In the next section, we argue that realistic prospects for future neutrino telescopes are much more encouraging.


\section{Near-Future Discovery Prospects}
\label{sec:future}

\begin{figure}
    \centering
    \includegraphics[width=\columnwidth]{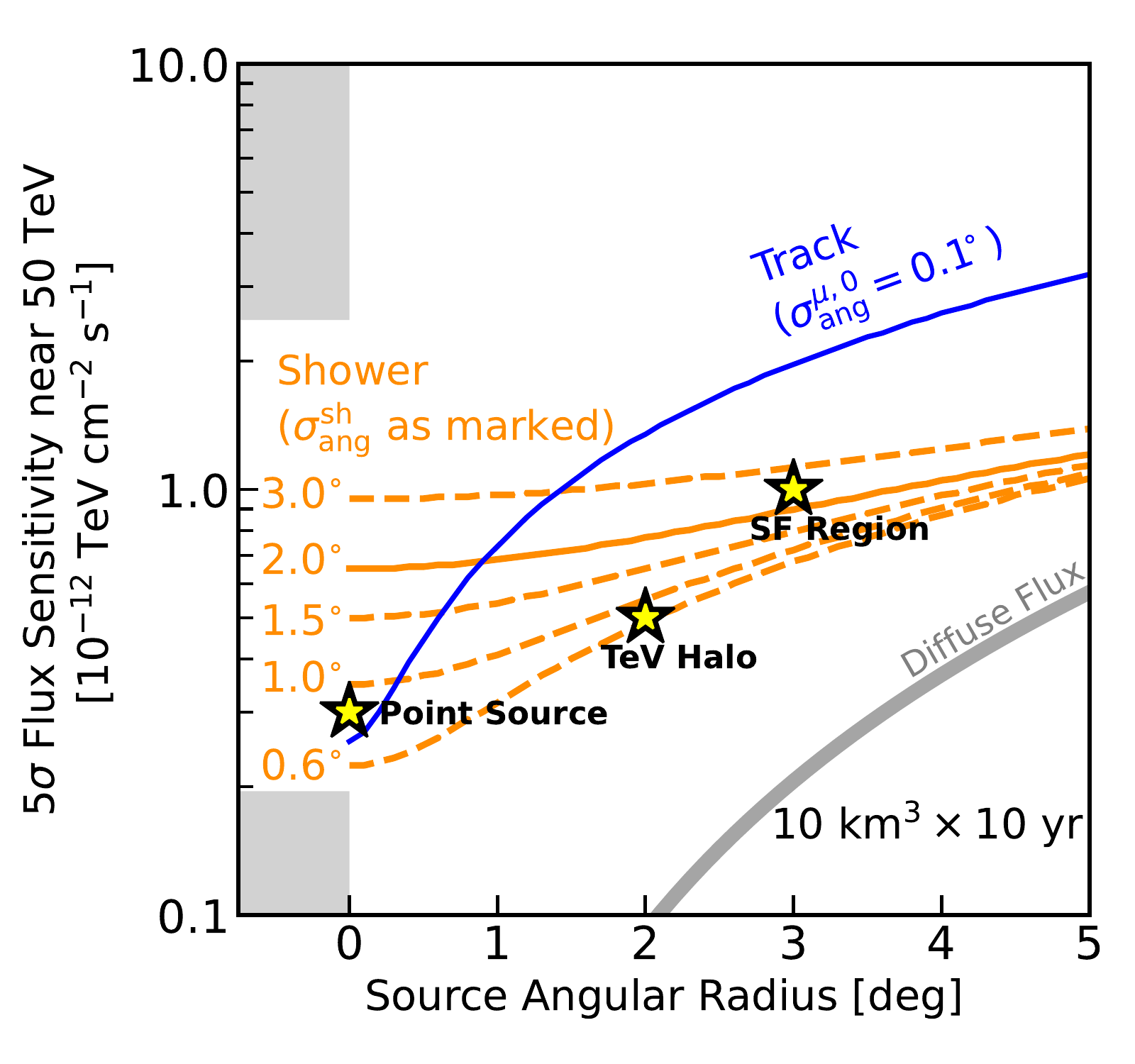}
    \caption{Same as Fig.~\ref{fig:main-sh-ang}, but for a larger volume of $V=10$~km$^3$, representing a network of future neutrino telescopes.  Stars indicate example sources.  The gray line shows the flux of isotropic astrophysical neutrinos. \emph{With the combination of track+shower detection and multiple neutrino telescopes, there are promising future prospects for discovering the Milky Way PeVatrons.}}
    \label{fig:near-future}
\end{figure}

In this section, we calculate the near-future prospects, which are better than shown in the previous section because future neutrino telescopes will be larger and there could be a network of them.  IceCube-Gen2~\cite{2021JPhG...48f0501A} (South Pole; $V \sim 8$~km$^3$) will likely be the largest. However, it seems very difficult to substantially improve the shower angular resolution for ice-based detectors, especially in the energy range of our interest, because it is largely determined by light scattering in the medium. 

Here, we instead focus on water-based detectors, which will have superior angular resolution. Near-future water experiments include: KM3NeT~\cite{2016JPhG...43h4001A} (under construction in the Mediterranean Sea; $V \sim 1 $~km$^3$), Baikal-GVD~\cite{2021chep.confE.606S} (under construction in the Lake Baikal in Russia;  $V \sim 1 $~km$^3$), the Pacific Ocean Neutrino Experiment (P-ONE)~\cite{2020NatAs...4..913A} (under planning and prototyping in the Pacific Ocean;  $V \sim 3 $~km$^3$), and the Tropical Deep-sea Neutrino Telescope (TRIDENT)~\cite{2022arXiv220704519Y} (proposed in the South China Sea;  $V \sim 8 $~km$^3$). Combining these detectors would effectively enhance the detector volume. Counting only water-based detectors, the total volume could be more than 10~km$^3$. For showers and contained-vertex muon tracks, as both event rates from source and background increase as $\propto V$, the TS increases approximately as $\propto \sqrt{V}$ (and similarly for the exposure time). For through-going muon tracks, the scaling of the TS is $\propto \sqrt{A}$.

Figure~\ref{fig:near-future} shows the flux sensitivities for a hypothetical combined 10~km$^3$ detector.  For channels that depend on the detector volume, combining detectors is linear.  For channels that depend on the detector area, like through-going muons, one would have to be more careful, though we neglect this because we assume that one potential water-based detector (TRIDENT) dominates the collection. Thus, the detector area is estimated as $10^{2/3}$~km$^2$. Other aspects of the calculation setup follow those in the previous section. Note that we assume a fixed detector latitude and a source visibility of 100\%; in fact, with multiple detectors located at different latitudes, some may have poor visibility for given sources. This needs to be compensated by a longer exposure time or larger detector volume.

For a few types of sources, we indicate representative sizes and fluxes by yellow star symbols:

\begin{itemize}

    \item \emph{Point Sources.} We show a point source that has a gamma-ray flux as large as the Crab flux near 100~TeV. For a point source, tracks would perform better; they can detect this hypothetical source, while showers are short of detection by a factor of $\sim$3. We note that the point-source emission from Crab (among many other gamma-ray point sources) is mostly due to leptonic emission. If such sources also accelerate hadrons, their neutrino emission is likely significantly more extended than observed in gamma rays, because the cooling time for hadrons is significantly longer than that of leptons.
    
    \item \emph{TeV Halos.} As an example for the ``vicinity of an accelerator" case, we show the Geminga TeV halo.  (Again, Geminga is most likely leptonic, but we use its flux to set a scale.) We extrapolate the gamma-ray flux from $E_\gamma=7$~TeV to 100~TeV, assuming an $E_\gamma^{-2.23}$ power-law, as observed by HAWC with a 2$^\circ$ template~\cite{2017ApJ...843...40A}. Using the extended source HESS J1825-137, which might be in transition stage between a pulsar wind nebula to a TeV halo, would yield a similar flux.  Showers can detect sources somewhat brighter than the Geminga halo. 
    
    \item \emph{Star-Forming Regions.} As discussed above, regions of active star formation and dense gas may produce a neutrino flux of $\sim$10$^{-12}$~TeV~cm$^{-2}$~s$^{-1}$ over a few-degree scale. Showers could detect such emission with a modest angular resolution of $\sim2^\circ$.
    
\end{itemize}

The above examples suggest that the shower channel is promising for detecting realistic extended sources in the Milky Way. Our results motivate new shower analysis that combines multiple water-based detectors to effectively increase the detector volume. While we have only compared fluxes near 50~TeV, the detectability weakly depends on the source spectrum. Our assumed spectrum, given in Eq.~(\ref{eq:neutrino}), lies between two commonly assumed cases, $E_\nu^{-2}$ and $E_\nu^{-3}$.

When observing extended sources, the background can be large. In particular, the diffuse \emph{astrophysical} neutrinos (both Galactic + extragalactic) may start to be important for high energies. For reference, in Fig.~\ref{fig:near-future} we show the astrophysical background flux (integrated over the putative source angular radius and taking angular resolution into account) at 50~TeV. Over the source sizes we consider, this is not a significant background.

Finally, we note that the dependence of sensitivity on the detector volume is modest, $\propto \sqrt{V}$, while increasing the shower angular resolution makes the sensitivity better $\propto 1/\sigma_{\rm ang}^{\rm sh}$. This suggests that for detecting Milky Way PeVatrons, it may be better to make a detector more densely instrumented than bigger. In this regard, TRIDENT appears promising due its varying spacing of the optical modules plus a proposed new type of module that enhances the angular resolution~\cite{2022icrc.confE1043H}.


\section{Conclusions}
\label{sec:conclusion}

In the Milky Way, there exist unidentified natural particle accelerators that produce the observed bright diffuse flux of TeV--PeV hadronic CRs. While TeV--PeV gamma-ray observations have significantly improved our understanding of the possibilities for these sources, they remain mysterious.  Decisively separating hadronic from leptonic sources requires detecting neutrinos, which are only produced in hadronic sources.  This has not been yet achieved, despite more than a decade of searches by IceCube and ANTARES.  

In Ref.~\cite{2023PhRvD.107d3002S}, we established a new multi-messenger framework designed to provide a systematic approach to understanding the Milky Way's hadronic PeVatrons.  As part of this, we delineated possibilities for why neutrino sources have not yet been detected.  In this paper, we detail one of those possibilities, that PeVatrons have escaped detection due to large source angular sizes.  Indeed, gamma-ray observations, especially with WCDs, have discovered a wide variety of extended sources, ranging from halos around likely CR accelerators to clusters of sources (see Fig.~\ref{fig:schematic}).

In this paper, we show that the neutrino-induced  shower channel has advantages for detecting extended sources compared to the commonly used neutrino-induced track channel.  This is due to the much lower atmospheric background intensity for the shower channel.  While IceCube has poor angular resolution for showers, this is due to light scattering in ice, and water-based detectors are expected to have much better resolution.  More generally, we emphasize that the shower channel has much less {\it intrinsic} limitations than the track channel, even if it has more {\it technical} limitations at present.  In other words, the shower channel has much more room to improve with dedicated efforts.

Among our quantitative results, we calculate the unique potential of showers to detect emission from hadronic PeVatrons.  We show that if good angular resolution can be achieved, then showers are a more powerful technique than tracks for detecting extended sources.  However, building on our results in Ref.~\cite{2023PhRvD.107d3002S}, we find that detectors bigger than 1~km$^3$ will almost certainly be needed for detecting point or extended sources.  Individual detectors, or a collection thereof, reaching the 10~km$^3$ scale will likely be needed.  As noted above, improving detector angular resolution can be as important as increasing detector mass.

Finding Milky Way's hadronic PeVatrons is a dream in neutrino astrophysics, and a comprehensive approach is needed.  This dream is encouraged by the bright diffuse flux of hadronic CRs and the many bright point and extended sources seen in gamma rays.  It may be that future detectors will soon find the neutrino sources via the track channel. However, it may instead be that PeV CRs escape the acceleration sites with little hadronic interaction, so that they are not visible with gamma rays and neutrinos. In such a pessimistic scenario, showers can help to detect extended emission from diffusing CRs. In any case, the importance of finding the sources of the CRs motivates bold actions.


\section*{Acknowledgments}

We are grateful for helpful discussions with Ivan Esteban, Darren Grant, Antoine Kouchner, Stephan Meighen-Berger, Sergio Palomares-Ruiz, Mehr Un Nisa, Donglian Xu, Bei Zhou, and especially Francis Halzen and William Luszczak. This research made use of {\sc matplotlib}~\cite{matplotlib} and {\sc numpy}~\cite{numpy}.  

T.S.\ was primarily supported by an Overseas Research Fellowship from the Japan Society for the Promotion of Science (JSPS) and a JSPS PD Research Fellowship.  T.S.\ was partially supported by and J.F.B.\ was fully supported by National Science Foundation Grant No.\ PHY-2012955.

\newpage
\clearpage

\appendix
\clearpage


\section{through-going Muon Spectra}
\label{app:throughgoing}

The CC interactions of muon neutrinos with nucleons produce muons, which leave track-like signatures in the detector. Due to the long distances they can travel, those produced outside the detector can still reach the detector, observed as through-going events. For a neutrino source with a flux of $F_{\nu_\mu} = d^3N_{\nu_\mu}/(dtdAdE_\nu)$, we approximate the observed spectrum of through-going events as 

\begin{widetext}
\begin{equation}
\begin{aligned}
\frac{dN_\mu}{dE_\mu^{\rm fin}} = TA
\int_{\frac{E_\mu^{\rm fin}}{1-y_{\rm \scriptscriptstyle CC}}}^\infty dE_\nu
\int_0^1 dy_{\rm \scriptscriptstyle CC} 
\int_0^{X_\oplus(\theta_Z)} dX \, N_A\sigma_{\rm \scriptscriptstyle CC}(E_\nu)
F_{\nu_\mu}(E_\nu)e^{-\tau(E_\nu, X)}
\frac{1}{\sigma_{\rm \scriptscriptstyle CC}}\frac{d\sigma_{\rm \scriptscriptstyle CC}}{dy_{\rm \scriptscriptstyle CC}}(E_\nu, y_{\rm \scriptscriptstyle CC})
\frac{dP}{dE_\mu^{\rm fin}}(E_\mu^{\rm init}, E_\mu^{\rm fin}, X),
\end{aligned}
\label{eq:mu_full}
\end{equation}
\end{widetext}

where each component is as follows:
\begin{itemize}
 
    \item $T$ is the observation time. 
    
    \item $A$ is the detector physical area. Note that this is different from the effective area, $A_{\rm eff}$, which accounts for several factors in Eq.~(\ref{eq:mu_full}) as well as detector efficiency following search cuts.
    
    \item $y_{\rm \scriptscriptstyle CC}$ is the inelasticity in CC interactions. The energy of muons at production is related to the parent neutrino energy as $E_\mu^{\rm init} = (1-y_{\rm \scriptscriptstyle CC})E_\nu$. The distribution of $y_{\rm \scriptscriptstyle CC}$ at muon production is accounted for with the normalized differential cross section, $(\sigma_{\rm \scriptscriptstyle CC})^{-1}d\sigma_{\rm \scriptscriptstyle CC}/dy_{\rm \scriptscriptstyle CC}$. 
    
    \item We define the distances (more precisely, column densities) as follows.  For a neutrino coming from a particular direction, the maximum possible distance through Earth to the detector is $X_\oplus$, which depends on the source zenith angle, $\theta_Z$. Then, if the neutrino interacts at an intermediate point, the distance traveled by the neutrino is $X_\oplus - X$ and the residual distance traveled by the muon is $X$. We calculate $X_\oplus$ as a function of $\theta_Z$ based on the Earth model of Ref.~\cite{1996APh.....5...81G}.
    
    \item $N_A = 6.0\times10^{23}$ is the Avogadro number in unit of g$^{-1}$, i.e., the number of nucleons per gram. 
    
    \item The optical depth $\tau = N_A(\sigma_{\rm \scriptscriptstyle NC} + \sigma_{\rm \scriptscriptstyle CC})(X_\oplus - X)$ accounts for the neutrino absorption by Earth after propagating over $X_\oplus - X$, where $\sigma_{\rm \scriptscriptstyle NC}$ is the neutral-current (NC) cross section. We regard neutrinos that experienced NC interactions as absorbed and do not consider their contribution, as substantial fraction of their energy is lost to hadrons at each interaction. We define $\tau_\oplus(E_\nu) = \tau(E_\nu, X=0)$, which accounts for the absorption all the way down to the detector.
    
    \item ${dP}/{dE_\mu^{\rm fin}}(E_\mu^{\rm init}, E_\mu^{\rm fin}, X)$ is the probability density that a muon with an initial energy of $E_\mu^{\rm init}$ reaches the detector with an energy of $E_\mu^{\rm fin}$ after traveling a column density $X$. This probability is zero for $E_\mu^{\rm fin} > E_\mu^{\rm init}$, which sets the lower limit on the integral in $E_\nu$.

\end{itemize}

This is reduced to a simple form with two approximations, treating the muon-energy loss and differential cross-section distributions as delta functions.  First, the muon energy losses are approximated by their mean value,
\begin{equation}
    \frac{dE_\mu}{dX} = -\alpha_\mu - \beta_\mu E_\mu,
    \label{eq:mu_loss_avg}
\end{equation}
so that
\begin{equation}
    \frac{dP}{dE_\mu^{\rm fin}} = \frac{\delta^D(X-X_0)}{\alpha_\mu + \beta_\mu E_\mu^{\rm fin}},
    \label{eq:mu_appro_X}
\end{equation}
where $\delta^D$ is the Dirac delta function and
\begin{equation}
    X_0 = \frac{1}{\beta_\mu}\ln{\frac{\alpha_\mu + \beta_\mu E_\mu^{\rm init}}{\alpha_\mu + \beta_\mu E_\mu^{\rm fin}}}.
\end{equation}
Second, the distribution of the $y_{\rm \scriptscriptstyle CC}$ 
is approximated as:
\begin{equation}
    \frac{1}{\sigma_{\rm \scriptscriptstyle CC}}\frac{d\sigma_{\rm \scriptscriptstyle CC}}{dy_{\rm \scriptscriptstyle CC}} = \delta^D(y_{\rm \scriptscriptstyle CC}-\langle y_{\rm \scriptscriptstyle CC} \rangle),
    \label{eq:mu_appro_y}
\end{equation}
Using Eqs (\ref{eq:mu_appro_X}) and (\ref{eq:mu_appro_y}), and $X_\oplus \gg X_0$, we obtain 
\begin{equation}
\begin{split}
    \frac{dN_\mu}{dE_\mu^{\rm fin}} = &\frac{TAN_A}{\alpha_\mu + \beta_\mu E_\mu^{\rm fin}} \\
    &\times\int_{\frac{E_\mu^{\rm fin}}{1-\langle y_{\rm \scriptscriptstyle CC}\rangle}}^\infty dE_\nu \, \sigma_{\rm \scriptscriptstyle CC}(E_\nu)F_{\nu_\mu}(E_\nu) e^{-\tau_\oplus(E_\nu)}.
\end{split}
\label{eq:ana_going}
\end{equation}
As we identify the observed muon energy as the detectable energy for the detector, we set $E_{\rm det} = E_\mu^{\rm fin}$. Throughout the main text, we use Eq.~(\ref{eq:ana_going}). We have used the publicly available code PROPOSAL~\cite{2013CoPhC.184.2070K} to verify that properly integrating over the two distributions (as in Eq.~\ref{eq:mu_full}) changes the resulting event rate only by less than $\sim$20\%.

Note that the mass density of the material in which the muons are produced, $\rho_m$ [g~cm$^{-3}$], does not appear in Eq.~(\ref{eq:ana_going}), which can be understood as follows. If a muon travels a distance of $L_\mu$ [cm] in a material which has the number density of nucleon of $n_n$ [cm$^{-3}$], then the interaction probability is $P\sim L_\mu n_n\sigma_{\rm \scriptscriptstyle CC} = (\rho_m L_\mu)(n_n/\rho_m)\sigma_{\rm \scriptscriptstyle CC}$. By definition, $X=\rho_m L_\mu$ and $n_n = N_A\rho_m$ (note that $N_A$ defined as ``nucleons per gram" is approximately the inverse of the nucleon mass, which is independent of the type of target material), and thus we reach $P=X N_A \sigma_{\rm \scriptscriptstyle CC}$, meaning that the $\rho_m$ factor is hidden. In other words, muons passing through a material with higher $\rho_m$ travel less distance $L_\mu$ (due to increased grammage in a given distance, and hence energy loss) but have an increased chance of CC interaction in a given distance (due to a larger $n_n$). 

In the energy range of interest, muon radiative losses are dominant over ionization losses. In the main text, we take $\beta_\mu = 3 \times 10^{-6}$~cm$^2$~g$^{-1}$, which is calculated for water in Ref.~\cite{Groom:2001kq}. If we instead choose the results for standard rock, the value for $\beta_\mu$ is by a factor of 1.3 larger, resulting in a larger muon energy loss (and hence reduced event rate). Note that this only takes into account the chemical composition, and the mass density difference between water and rock does not appear, as discussed above. 


\section{Choice of Track Angular Resolution}
\label{app:mu_ang}

\begin{figure}[b]
    \centering
    \includegraphics[width=\columnwidth]{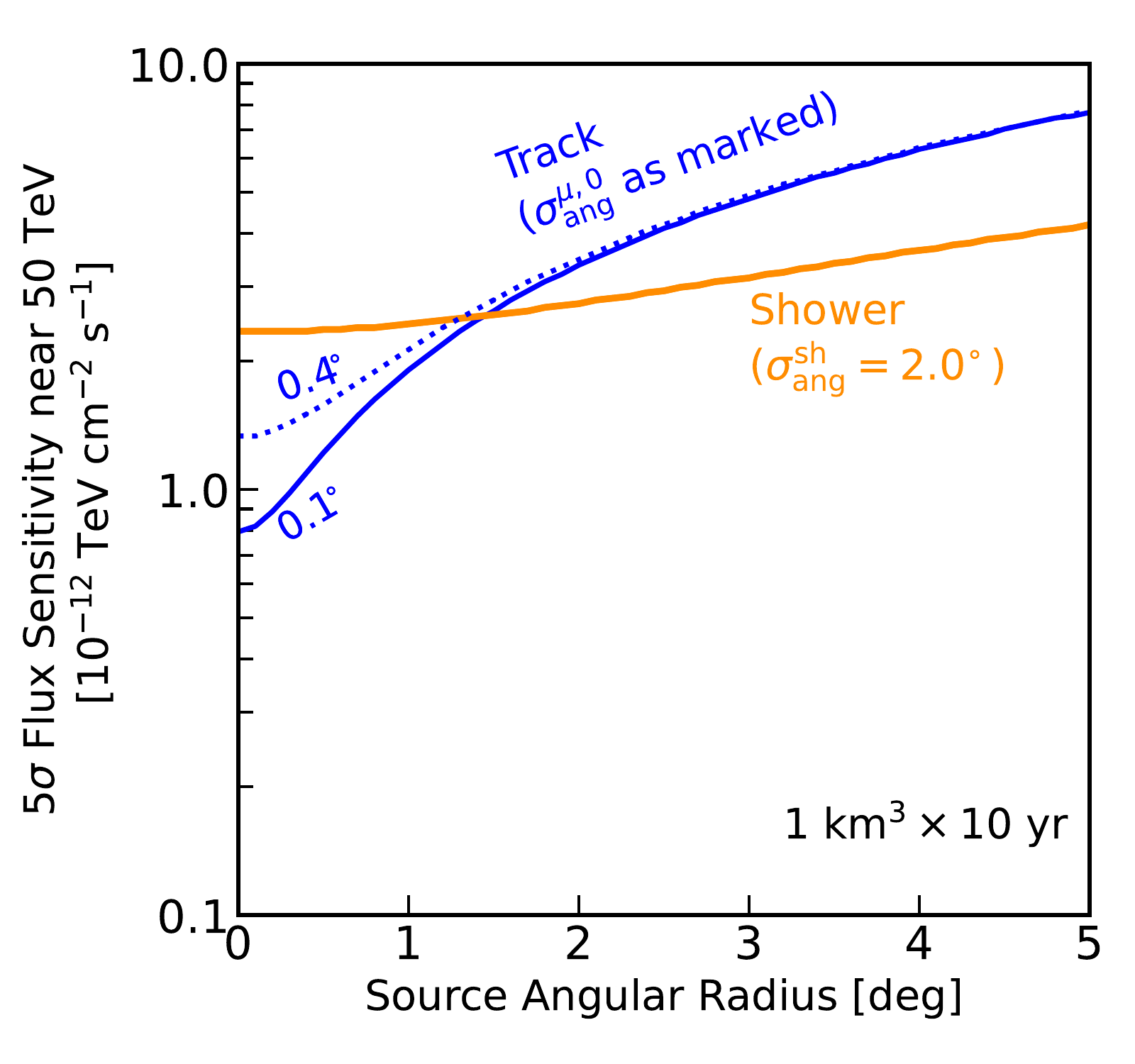}
    \caption{The same as Fig.~\ref{fig:main}, but we include the case where track angular resolution is 0.4$^\circ$ (blue dotted line) and omit the color shading.}
    \label{fig:mu-ang-04}
\end{figure}

Figure~\ref{fig:mu-ang-04} shows the case where the track angular resolution is worsened to 0.4$^\circ$ (blue dotted line), along with the fiducial case of 0.1$^\circ$ (blue solid).  The implications are discussed in the main text.

\newpage
\clearpage

\bibliography{icecube_source}

\end{document}